# Investigating the role of Cu foil orientation in the growth of large BN films synthesized by reactive RF magnetron sputtering

Nilanjan Basu[a], Alapan Dutta[b], Ranveer Singh[b], Tapobrata Som[b], Jayeeta Lahiri[a*].

[a] School of Physics, University of Hyderabad, Hyderabad, India-500046.
[b] Institute of Physics, Bhubaneswar, India -751005.

* Corresponding author: jlsp@uohyd.ac.in

**Abstract:** Two dimensional materials are an emerging class of materials which is transforming the present day research activity on a phenomenal scale. Hexagonal boron nitride is a wide band gap 2D material which is an excellent substrate for graphene based electronics. To achieve the full potential of hBN scalable and high yield growth procedures are required. Here, we demonstrate the synthesis of hBN by reactive R.F magnetron sputtering over copper foil. Copper foil preparation conditions determines the phase selectivity of BN films. Deposition of hBN on non-electropolished Cu foils with predominant (100) orientation resulted in growth of BN islands with mixed cubic and hexagonal BN phase. On electropolished Cu foils with high symmetry hexagonal (111) surface termination we get growth of continuous hexagonal BN films, while on Cu foils having (100) and (110) orientation with lower symmetry growth of cubic BN films are observed.

**Keywords**: cubic and hexagonal Boron Nitride, Cu foil, RF sputtering, EBSD, XPS

## 1. Introduction

Boron Nitride (BN) is wide band gap semiconductor ($E_g \sim 6eV$) with exceptional thermochemical stability, structural, electronic, and thermal conductivity properties [1]. BN has different polymorphs such as graphite like hexagonal (hBN), rhombohedral (rBN), diamond like cubic (cBN) and wurtzite (wBN) [2]. Most of the research is focused primarily on Cubic BN and hexagonal BN polymorphs since they have tremendous scope in different applications. hBN has a layered structure similar to graphite with an in-plane lattice mismatch of 1.7% , however in the out of plane direction it has an AA type of stacking where each B atom sits on top of N atom in the next layer [3]'[4]. hBN with its atomically smooth surface and homogeneous charge distribution became popular as the preferred dielectric substrate for graphene based devices [5]·[6]·[7]. Since then it has firmly established its place as one of the member of the expanding 2D materials family. It is now frequently used as tunnel barriers and encapsulation layers for nanoelectronic devices[8]·[9]·[10]. hBN in its new avatar has caught the imagination of researchers worldwide, which has led to unveiling of new properties. It has a thickness dependent photoluminescence property, with bulk hBN exhibiting high exciton luminescence in the deep UV region making it suitable for deep-UV photonics applications[11]·[12]. It is also a hyperbolic metamaterial with dielectric constant same in the basal plane but having opposite signs in normal plane[13]. Recently monolayer and few layer hBN films have exhibited promising response in resistive switching based memory devices with lower switching voltage, faster response time and longer retention [14]·[15]. It is also being used to fabricate planar and vertical heterostructures with other 2D materials for a plethora of applications [16]·[17]. For most of this proof of concept experiments exfoliated hBN is primarily used. To harness the properties of hBN optimally and to utilize it for speific applications, there is an immense need to produce large area high quality films in a scalable manner.

Large area hexagonal boron nitride thin films can be synthesized by number of methods- like chemical vapor deposition (CVD), molecular beam epitaxy (MBE) and radio frequency (RF) sputtering[18]. The crystallinity and crystal orientation of the substrate plays a crucial role in determining the quality of the hBN films. CVD is the most popular method to synthesize large area 2D materials including hBN [19]·[20]. CVD synthesis employs different kinds of transition metal substrates for growth of 2D materials[21]·[22]·[23]. Copper foils are the most popular choice as substrate for synthesis of hBN since they are relatively inexpensive and the carbon solubility in copper is also low[24]·[25]. However in CVD it is difficult to get repeatable growth rates and high processing temperatures are required to produce high quality films. Physical Vapor Deposition method like RF sputtering can also be used to synthesize uniform films of 2D materials over large areas [26].With RF magnetron sputtering we can deposit hBN films at a lower temperature with precise control over the growth rate and it is also scalable in nature.

Unlike CVD where hBN films are synthesized on metallic substrates which acts as catalyst, with RF sputtering hBN films can be deposited on noncatalytic substrates also[27]. RF sputtering has been extensively used in the synthesis of cubic BN films where cBN films mixed with amorphous, hexagonal phases were mostly reported[28]. In this paper, we report the growth of ultrathin hBN films on Cu foils using reactive RF magnetron sputtering. We investigated how the Cu foil surface termination plays a decisive role on the phase selection of BN film. We employed cyclic deposition at low temperatures followed by annealing to synthesize multilayer hBN films. Our study will be useful for synthesizing homogeneous uniform films of large area hBN on copper for both research and industrial purpose.

## 2. Experimental Method

### 2.1 Cu foil preparation

The as received copper foil (25 μm thick, 99.8% purity sourced from Alfa Aesar) was first cleaned in dilute nitric[29] acid for about 20s to remove the native oxide from its surface. It was then rinsed with deionized water followed by cleaning with acetone and isopropyl alcohol for removing any organic contaminants. Different methods were followed to further process the Cu foil which are described in the table below

Table 1: Cu foil processing conditions

| Method I | The non-electropolished Cu foil was annealed at 900°C in Ar gas for 120 min. |
| Method II | The electropolished Cu foils were annealed at 900°C in Ar:$H_2$ (90:10) gas for 120 min. |
| Method III | The electropolished Cu foils were annealed at 1040°C in Ar:$H_2$ (90:10) gas for 60 min. |

In the text, Cu foils processed using these methods are labelled as ~~Cu~~ M I, ~~Cu~~ M II and ~~Cu~~ M III respectively. Boron nitride films were synthesized on these copper foil with reactive radio frequency (RF) magnetron sputtering (25 W RF power, hBN target -2 inch diameter with 99.99% purity) in high purity

Ar:$N_2$ (10:1) gas mixtures with total pressure $10^{-2}$ mBar. For all depositions these parameters were kept constant. During growth the deposition time (t) was varied from 30 min to 3 hours while substrate temperature ($T_s$) was varied from RT to 400°C respectively. After deposition the substrate was annealed at 800°C for 40 min. Single step and cyclic/multiple step (deposition alternated by post annealing at 800°C for 40 min) deposition were used for synthesizing hBN films.

## 2.2 Characterization methods

The crystallographic orientation of the copper foil was studied by X ray diffraction using XRD, PAN analytical:X'Pert[3] Powder with Cu Kα line (Kα1 and Kα2 with 2:1 ratio). Electron back scattered diffraction (EBSD) using AMETEK EBSD detector fitted to FEI NOVA NANO 460 FESEM, the imaging was done at 20 KeV, sample tilted at 70°). X-ray photoelectron spectroscopy was recorded using Axis Ultra system equipped with monochromatic Al Kα X-ray source (1486.6 eV). The energy resolution was set at 0.1 eV. For Raman and AFM measurements, the hBN film was transferred to $SiO_2$/Si substrate using wet transfer method[30]. Atomic force microscopy (Asylum Research) was applied for AFM measurements. Raman spectra was acquired using Witec Alpha 300 spectrometer with 532 nm laser, 100X objective (675 nm laser spot) and 1mW input power. Transmission Electron microscopy (TEM) analysis was done using a FEI Tecnai $G^2$ 20 TWIN transmission electron microscope and has been used to obtain the SAED pattern. The aperture was 20 µm and accelerating voltage was 200 kV. The film from copper substrate was transferred over a copper grid with carbon nanomesh by wet transfer method which has been applied in previous transfers. FESEM images were acquired using a Carl Zeiss Ultra 55 model operated at 5kV. Conductive atomic force microscopy (cAFM) measurements were carried out over the h BN/Cu film. Asylum Research cAFM has been used for the I-V measurements by using a Pt/Ir conductive tip with compliance current of 20 nA.

## 3. Results and discussion

The as received cold rolled Cu foil is polycrystalline in nature with a dominant (200) texture. The presence of rolling lines in the as received Cu foil makes it very rough (roughness ~ 96 nm) (Figure S1). To improve the surface texture and to reduce the roughness of the Cu foil, it was subjected to additional processing as mentioned in the Materials and Methods section. Cu foils prepared by Method I, still had mixed texture, however the average grain size had increased to ~50 μm and the roughness has also decreased to ~16 nm. (Figure S1). From XRD measurements we observe the film has preferential (200) orientation.

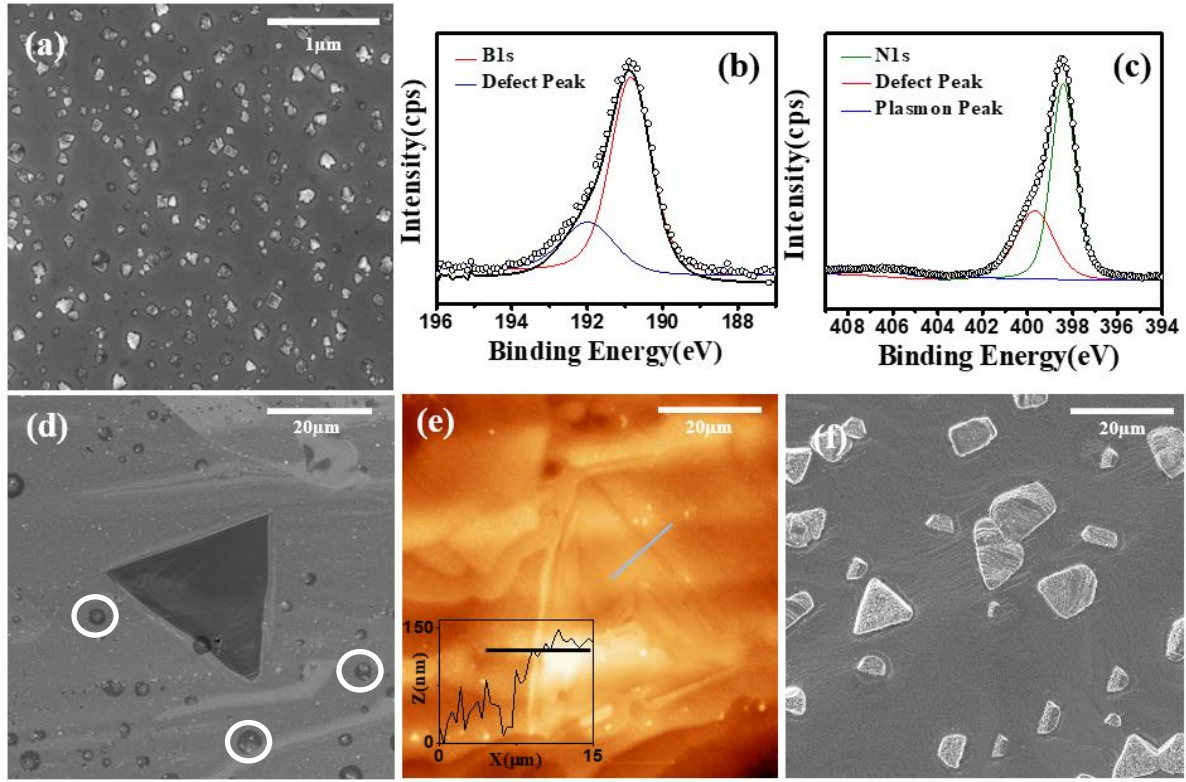

**Figure 1: hBN deposited on non-electropolished Cu substrate M I** (a) FESEM micrographs of hBN grains deposited at $T_s$ = RT for 60 min. (b) and (c) High resolution XPS spectra of B1s and N1s of the sample in panel (a). (d) FESEM micrographs of hBN grains deposited at RT using two step deposition ($t_1$=30min, $t_2$= 30 min). (e) AFM

topography image of the same hBN island on copper. Inset shows the height profile of the BN islands. (f) FESEM micrographs of hBN deposited at $T_s = 400°C$ using two step deposition ($t_1$=30min, $t_2$=30 min).

### 3.1 hBN films on non-electropolished Cu foil

The hBN film synthesized on Cu substrate MI at RT for 60 min is very discontinuous and with small grains nucleated all over the substrate (Figure 1(a)). Figure 1(b) and (c) shows the high resolution XPS B1s and N1s spectra of these films. From the fitting we can deconvolute the N1s (B1s) peak into 2 components at 398.4 eV (190.8 eV) and 399.6 eV (191.9 eV). The low energy peaks correspond to multilayer hBN films and the peak positions are similar to that reported for multilayer hBN films synthesized on Cu foils using CVD [31] . The peaks at higher binding energy can be attributed to defects or cubic BN in the deposited hBN film[31]. In the N1s spectra we also observe the characteristic $\pi$ plasmon loss peak at 406.2 eV associated with sp$^2$ bonded materials. Figure 1(d) shows FESEM micrograph of hBN synthesized on Cu substrate (MI) for total 60 min in two steps ($t_1$=30 min, $t_2$=30 min) at substrate temperature $T_s$= RT. We observe large triangular grains with edges 10-30 µm and some secondary nucleation in the remaining area. We also observe smaller islands with semi hexagonal shape (white circles). Triangular islands with N-terminated zig zag (ZZ) edges is the equilibrium shape of hBN on strongly interacting substrates like Ni(111), Ru(0001), Co(0001) [32],[33],[34]. The equilibrium shape of hBN vary from triangular (Nitrogen rich), truncated triangular to hexagonal (Boron rich) depending on the partial pressure of Nitrogen and Boron atoms [35]. In our growth condition we have higher Nitrogen partial pressure so the triangular shaped islands have grown at the cost of semi hexagonal shaped islands. The presence of these triangular shaped islands thus indicates the formation of hBN phase. Figure 1(e) shows the AFM image of the hBN triangular grain with 12 µm lateral dimensions and 130 nm tall. These discontinuous hBN islands imply that the BN adatoms are not wetting the Cu surface. Increasing the substrate temperature during deposition resulted in lower nucleation and higher wetting of the substrate, however the surface diffusion of hBN adatoms during post growth annealing is not high enough to overcome the high surface roughness. The discrete triangular hBN islands therefore grow in 3D manner. Figure 1(f) shows the FESEM micrograph of the BN film where we can see that at 400°C substrate temperature for total 60 min in a two-step growth ($t_1$=30min, $t_2$=30 min). We now observe a higher density of nucleated hBN islands with mixture of triangular and trapezoidal shape. On Ir(111) and Pt (111) substrates which have weak interaction with hBN (similar to Cu), the energetically stable shape of hBN islands also depend on the step morphology of the underlying substrate, trapezoidal shapes of hBN have also observed in addition to the triangular islands [36].

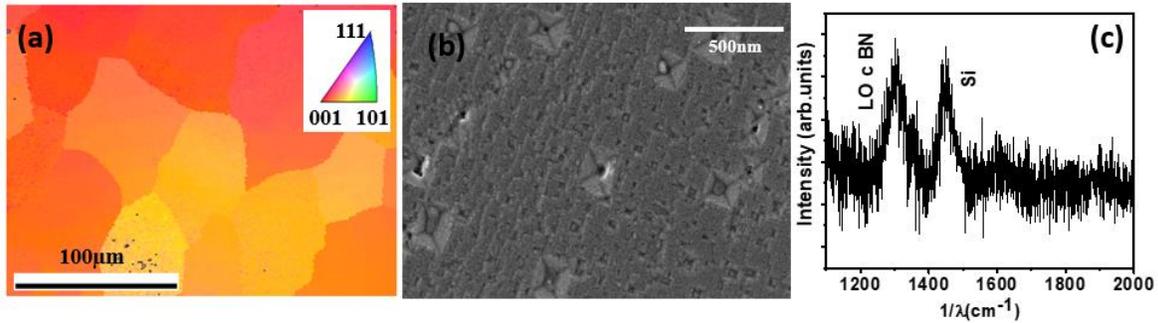

**Fig.2 BN deposited on electropolished Cu substrate MII** (a) EBSD map of the copper substrate, Inset-IPF legend. (b) FESEM micrograph of the BN film over copper deposited at $T_s$= 400°C using two step deposition ($t_1$=30min, $t_2$=30 min). (c) Raman spectrum of the deposited film after transferring.

### 3.2 hBN films on electropolished Cu foil

To improve the surface roughness the Cu foils were electropolished and then annealed at high temperature. The EBSD map of electropolished Cu foil prepared by MII showed that the Cu foil is comprised mostly of large 100 μm sized grains with (001) and (110) orientation (Figure 2(a)). In the XRD pattern we observe peaks for (100) and (110) planes (Figure S2). The foils also become smoother with surface roughness of the foil reduced to 10 nm. Figure 2(b) shows the FESEM micrograph of the BN film over the copper substrate. We can observe grains with truncated pyramidal shape in the film with square and quasi triangular facets. In carbon systems, diamond films with pyramidal micrometric crystals with square facets and triangular facets corresponding to {111} and {100} lattice plane have been observed on Si(100) wafers [37]. Similar grain growth with four quasi {111} faces was also observed for cBN films on Si(001) substrate synthesized by CVD [38]. From the morphology of the BN film we can conclude that the cubic BN phase is growing on Cu(001) surface. Raman spectrum of the film shown in Figure 2 (c) also corroborates our FESEM measurements. Even though Raman signals are very weak, we can see a clearly see a LO peak of cBN centered at 1303 cm$^{-1}$ and third order Si peak at 1450 cm$^{-1}$ [39]. FESEM micrographs and Raman spectroscopy of these films clearly indicate that cBN phase is formed on Cu (001) surface.

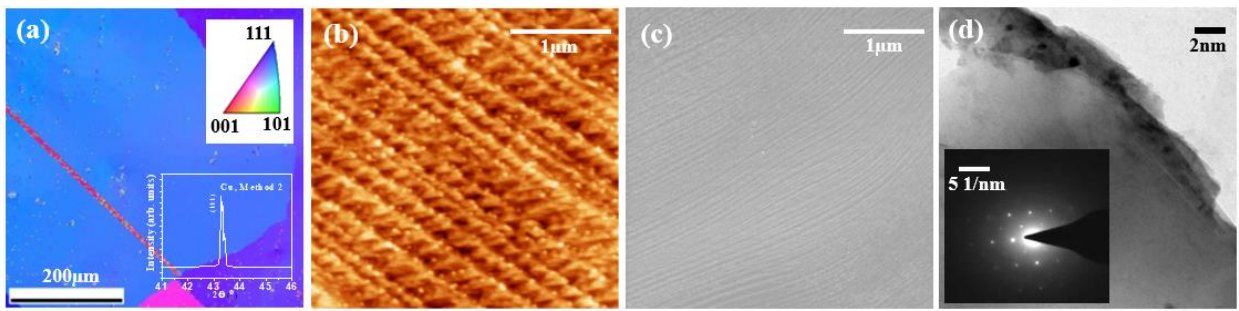

**Figure.3 – hBN deposited on electropolished Cu substrate MIII -** (a) EBSD orientation map of Cu substrate prepared by method 2. The inset shows the EBSD legend and XRD (b) AFM of the Cu substrate, (c) FESEM micrograph of hBN deposited at $T_s$ = 400°C using two step deposition (t$_1$=30min, t$_2$=30 min) (b) TEM image of hBN. Inset –SAED pattern.

Increasing the annealing temperature from 900°C to 1040°C in MIII caused a marked increase in the (111) surface termination as evident from the EBSD map of the surface in Figure 3(a). The Cu foil now has dominant (111) orientation with an average ~ 500 μm grain size. From AFM image in Figure 3(b) we can see the surface facets on the Cu foil and the surface roughness has further reduced to ~5 nm. Figure 3(c) shows the FESEM micrograph of the BN film deposited on these Cu substrates MIII at 400°C for total 60 min in two step cyclic deposition (t$_1$=30 min, t$_2$=30 min). The BN film synthesized is continuous and completely wets the substrate. Figure 3(d) shows the TEM image of the film after the film was transferred over copper grid with carbon coated nano mesh. The crystallinity of the deposited hBN was investigated by using electron diffraction in TEM. Figure 3(d) inset shows the selected area diffraction pattern (SAED) of the particular area of the film. We can clearly see the hexagonal spots in the SAED pattern which is in accordance with the reported [40] SAED pattern on of single crystalline hBN.

Table 2: Lattice constant of bulk Cu, cBN, hBN

| Cu | Lattice Constant (Å) | BN | Lattice Constant (Å) |
|---|---|---|---|
| Cu (bulk) | a = 3.59 | cBN (bulk) | a = 3.61 |
| Cu(001) | a = 3.59 (square) | hBN (bulk) | a = 2.50, c = 6.66 |
| Cu(110) | a = 3.59, b = 2.53 | --- | --- |
| Cu(111) | a = 2.53 (hexagonal) | --- | --- |

Cubic BN is the most stable phase at low temperature while hexagonal BN is the stable phase at ambient temperature and atmospheric pressure [41]. Since the difference in free energies of formation of c-BN and h-BN is very low, the lattice mismatch with the substrate determines the phase of BN films. The lattice mismatch between cBN and hBN with Cu(001) and Cu(111) surface is the lowest (Table 2). Hence on Cu(001) substrate cBN is the preferred phase while on Cu(111) the stable phase is hBN.

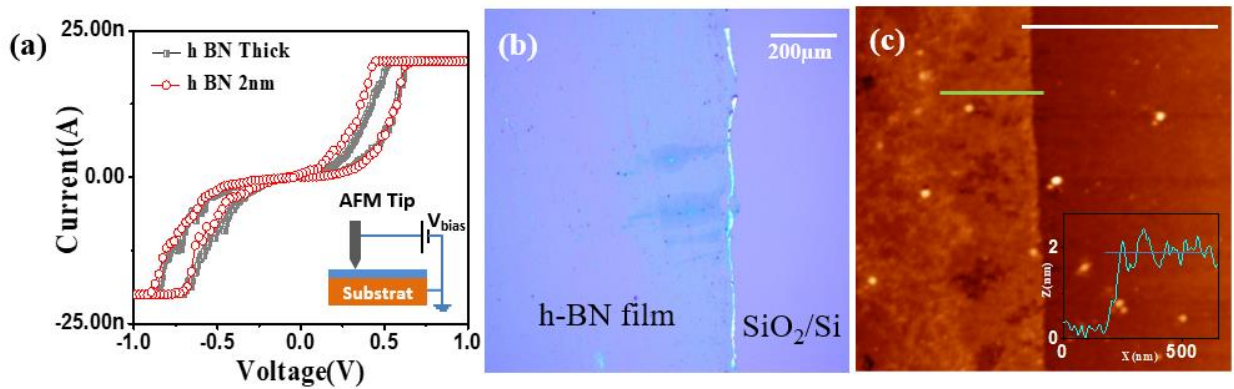

**Figure 4. Electrical property of hBN film** (a) I-V characteristic of hBN films of different thickness measured using CAFM. (b) AFM image of the transferred of hBN film transferred on $SiO_2$ (300 nm)/Si. This hBN film was synthesized using two step deposition for total 120 min at 400$^o$C Inset: Height profile of the film. Scale bar 1μm.

We investigated the electrical property of BN films prepared by M III by using cAFM. The current−voltage relationship (I−V) has been measured on hBN film synthesized for total of 120 min and 360 min respectively. A typical I−V characteristic recorded on Pt/hBN/Cu devices is shown in Figure 3(a). The I-V response of 2 nm and 7 nm thick film show a similar rectifying behavior. Figure 3(b) and (c) shows the optical micrograph and AFM image of the transferred hBN film on SiO2/Si substrate. AFM measurement of the film shows that its thickness is about 2 nm or 6 layers. After 3-4 cycles the hysteresis loop collapsed due to irreversible creation of defects.

## 4. Conclusion

We have successfully synthesized multilayer hBN films from 2 nm to 7 nm thick films (6-20 layers) over copper foil by radio frequency (RF) magnetron sputtering. We have demonstrated how Cu foil preparation plays a role in phase selection of BN films. On non-electropolished Cu foil with (001) surface termination we get discontinuous BN with mixed phase. On electropolishing the copper foil, cubic BN film grows on Cu (001) surface while on Cu (111) surface termination we get continuous uniform hBN films. This present work should have an impact on the realistic application of h BN by making it's synthesis more economic and thus pave the way for industrial scale production.

## Declaration of Competing Interest

The authors declare that no competing financial interests or personal relationships have influenced the work reported in this paper.

## Acknowledgement

This research was supported by DST SERB grant YSS/2015/000572. We would like to acknowledge Dr. Ravi Chandra Gundakaram, ARCI, Hyderabad and Dr. Jai Goutam, SEST, University of Hyderabad for EBSD measurements.   The authors acknowledge facilities provided by the Central Instrumentation Laboratory, University of Hyderabad. We would like to acknowledge Dr. Ravi Chandra

**Supplementary information:**

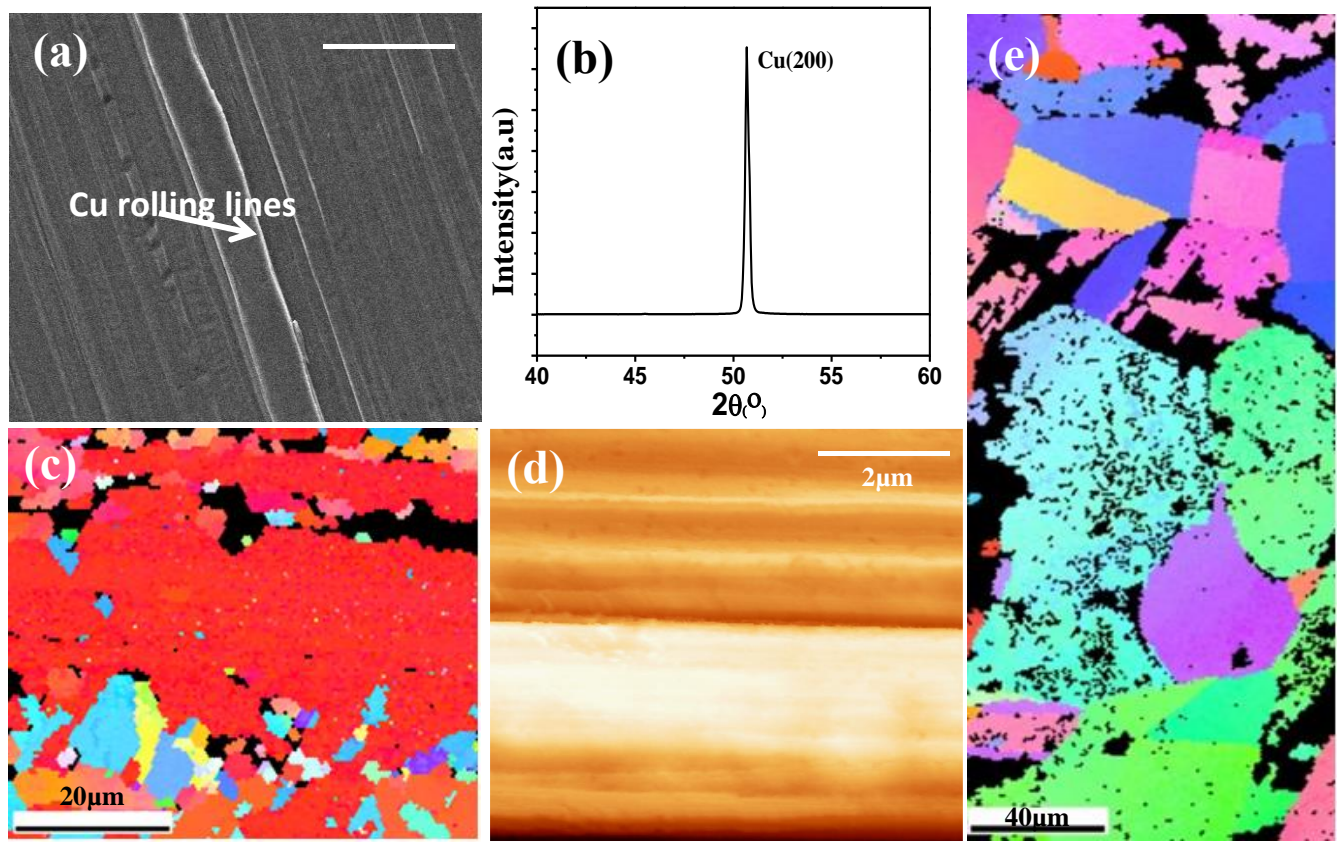

Figure.S1. (a) FESEM image of the as received copper foil. Rolling lines can be clearly seen which increases the r.m.s roughness. (b) XRD of the as received Cu foil with a dominant peak at 50.3° (Cu 200). (c) EBSD of the as received foil shows a polycrystalline texture with 5-6µm grain size. (d) AFM image of the as received foil, the r.ms. Roughness is 96 nm. (e) EBSD of the copper foil after treating it through method 1. The grain size has increased to 40-50 µm but it is still polycrystalline in nature.

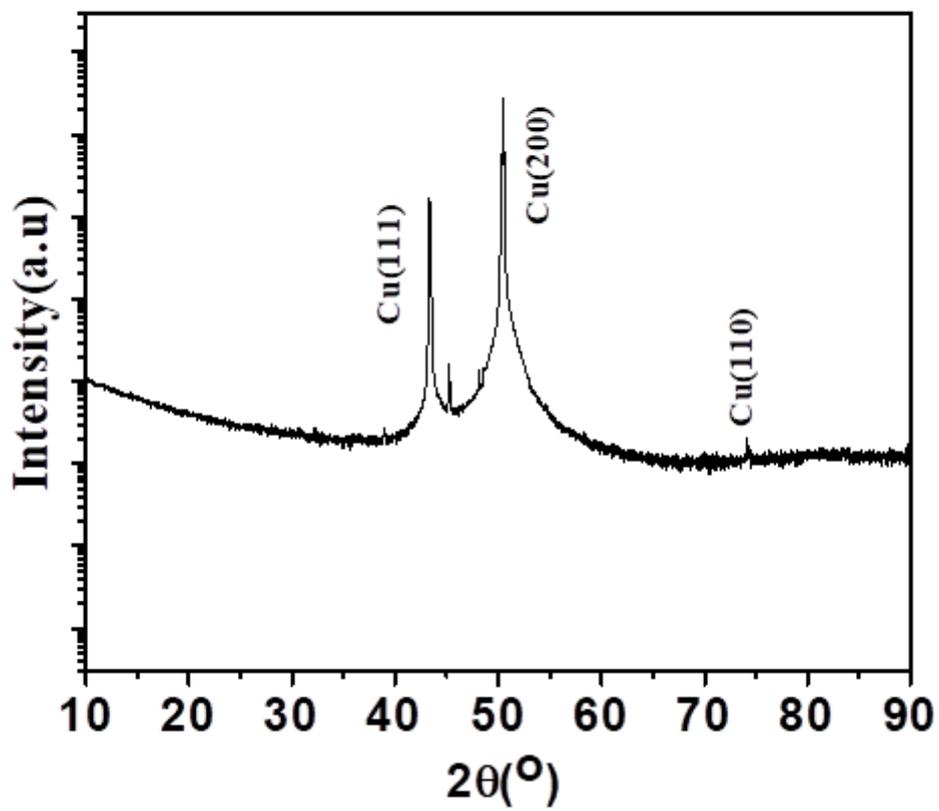

Figure.S2 – XRD of the copper substrate prepared by MII